# X-ray variability and 1mHz oscillations in TT ARI

A.Baykal[1,4], A.Esendemir[1], Ü.Kızıloğlu[1], M. A. Alpar[1], H. Ögelman[2], N.Ercan[3], and G.İkis[3]

[1] Physics Department, Middle East Technical University, Ankara 06531, Turkey
[2] Department of Physics, University of Wisconsin-Madison, 1150 University Ave., Madison, WI 53706, USA
[3] Department of Physics, Boğaziçi University, İstanbul, Turkey
[4] Laboratory for High Energy Astrophysics NASA/GSFC Code 666 Greenbelt, Marlyland 20771 USA

December 28, 1994

**Abstract.** Using the archival ROSAT observation of TT Ari, X-ray energy spectra in different orbital phases and power spectra of the intensity time series are presented. Spectral fits show that the source gets brighter during the observation. The orbital modulation of the X-ray counting rate and bremsstrahlung temperature suggests that soft X-ray emission peaks in the orbital phase interval 0.75-0.90, when an outer disk hot spot is near the line of sight. This correlates with the orbital modulation of C IV($\lambda$1549) absorption (Robinson and Cordova, 1993). Timing analysis indicates that while the source gets brighter, the frequency of the 1mHz oscillation is not correlated with X-ray intensity. This implies that in the X-rays from TT Ari, the beat frequency model is not appropriate for explaining the changes in the 1mHz oscillations.

## 1. Introduction

TT Arietis (BD +14°341) has been classified as a nova-like variable from photometric observations (Smak and Stepien 1969, Cowley et al. 1975). It was also found to be a hard X-ray source with the Einstein satellite (Córdova et al. 1981). The spectroscopic investigations (Cowley et al. 1975 and Thorstensen et al. 1985) indicated that TT Ari is a binary system with an orbital period of $0.^d13755$. However, this period is not consistent with the photometric period of the source (Smak and Stepien 1975 and Thorstensen et al. 1985). Detailed observations have shown that at the longest time scale TT Ari behaves as a VY Scl system ("anti-dwarf nova") with an intermediate polar (Hudec et al. 1984). It can remain at $V \sim 10.5$ (high state) for many years, interrupted irregularly by low states at $V \sim 16.5$, which last less than a year (Shafter et al. 1985). During high states, TT Ari has also shown rapid quasi-periodic oscillations up to 0.2 mag with periods between 14 and 27 min. (Semeniuk et al. 1987, Hollander and van Paradijs 1992). In the last three decades these oscillations have shown a secular decrease in period while the average optical luminosity and the inferred mass transfer rate roughly increased by 10% to 15%.

Recently Hollander and van Paradijs (1992) proposed that changes in the frequency of quasi-periodic oscillations in the optical may be explained by the beat-frequency model (Alpar and Shaham 1985). In this model small changes in the mass transfer rate can give relatively large changes in the quasi-periodic oscillation frequency. The beat frequency model fits the change in the quasi-periodic oscillation frequency, if the white dwarf has a rotation period of the order of a few hundred seconds.

Jensen et al. (1983) investigated the correlations of X-ray and optical flickering activity (or optical quasi-periodic oscillations) using simultaneous observations with the Einstein X-ray observatory, the Louisiana State University (LSU) Observatory and the Mount Wilson Observatory. They have found flickering activity in X-rays at $\sim 17$ min ($\sim 1$ mHz) timescales. They proposed that optical flickering in TT Ari is basically produced in the inner accretion disk, with a fraction of the energy transported to a different region where the X-ray flickering is produced.

In this work we examine the X-ray timing of TT Ari, and study its X-ray spectrum. In Section 2, we report the archival ROSAT observation of TT Ari. In Section 3, we present the X-ray spectra and the orbital modulation. In order to investigate the quasi-periodic oscillations we constructed the power spectra with various methods for unevenly sampled time series (Deeming 1975, Roberts et al. 1987). These are also discussed in Section 3.

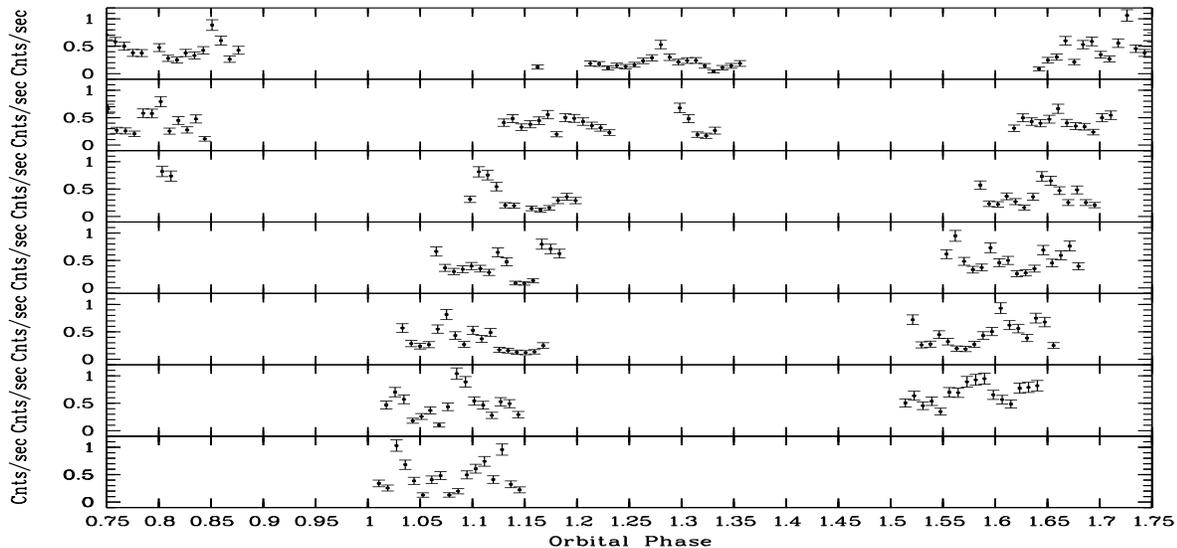

**Fig. 1.** The light curve of TT Ari with a start epoch TDB JD 2448469.8234 and binned with 100 seconds. The horiontal axis is the orbital phase based on the spectroscopic ephemeris (Thorstensen et al. 1985); the different panels show observations in chronological order. Each orbital cycle is thus given separately.

## 2. The Data

TT Ari was observed with the Position Sensitive Proportional Counter (PSPC) at the focus of the X-ray telescope of ROSAT. The PSPC is a gas filled proportional counter sensitive over the energy range 0.1-2.4 keV with an energy resolution $\Delta E/E \sim 0.43$ at 0.93 keV. Detailed descriptions of the satellite, X-ray mirrors, and detectors can be found in Trümper (1983) and Pfeffermann et al. (1986). The X-ray observations reported here were obtained between Aug. 1, 1991 (JD 2448469.8234) and Aug. 2, 1991 (JD2448470.7048) with a total effective exposure time of 25174 sec. The reduction of the ROSAT archival data has been performed with the EXSAS package (Zimmermann et al.1993). The TT Ari source counts were extracted from a circle of radius $2'.5$ which is expected to include 99% of the photons from the source, according to the point spread function of the PSPC. The background was determined from a source free area of radius $10'$. The mean background subtracted, vignetting and dead-time corrected count rate for the whole observation was $0.4129 \pm 0.0024$ counts s$^{-1}$. The light curve is presented in Figure 1. This light curve shows data for each individual orbital cycle separately.

## 3. Results

### X-Ray Spectra

The large number of counts obtained from TT Ari, about 11000, allow fits for various spectral models. Black-body, thermal bremsstrahlung, power law, and Raymond-Smith models were tried in the fits. All models give acceptable values of $\chi^2$. The results of our fits are listed in Table 1. In order to compare our results with the earlier results from the Einstein observatory (see the next section) we examine the thermal bremsstrahlung model. The best fit parameters of the thermal bremsstrahlung model in the 0.1 - 2.4 keV band give a column density $N_H = (4.7 \pm 0.2) \times 10^{20}$cm$^{-2}$, a temperature T = $5.1 \pm 1.4$ keV and an unabsorbed flux (extrapolated to 4.0 keV) $F_x = 11.3^{+0.3}_{-0.4} \times 10^{-12}$erg cm$^{-2}$s$^{-1}$ at the $1\sigma$ level. The implied source luminosity for a distance of 125 pc (Patterson 1984) is $L_x \sim 2.1 \times 10^{31}$erg s$^{-1}$. The thermal bremsstrahlung fit shows an excess which can be fitted by a Gaussian emission line at $0.996 \pm 0.062$keV, and can be improved further by including an absorption feature at $\sim 0.4$ keV to lead $\chi^2_\nu = 0.79$ per degree of freedom. The best spectral fit is shown in Figure 2.

### Orbital Modulation of X-ray Count Rates and Spectra

X-ray counts from the source region were folded with the spectroscopic ephemeris, with orbital period P=$0^d.13755114(13)$, given by Thorstensen et al. (1985), after correcting the photon arrival times to the solar system barycenter. The resulting orbital light curve is shown in Figure 3. ROSAT observations start at UTC 2448469.8234, i.e., $\sim$ 34465 orbits after the given ephemeris, the spectroscopic orbital period with the stated uncertainty produces an uncertainty in phase predictions of $\sim 0.0045$, which is negligible. The X-ray count rate is modulated at the $\sim 100\%$ level around the mean counting rate of $\sim 0.4$ counts s$^{-1}$. There were no observations in the phase intervals 0.4-0.5 and 0.9-1.0, due to ROSAT ob-servation windows. The count rate decreases from phase ~0.05 to ~0.35, is at a higher level at phase 0.55 and remains almost constant between phases ~0.55-0.90. To search for possible orbital modulation of energy spectra, the photons observed during the orbital phases 0.90-0.25, 0.25-0.40, 0.50-0.75 and 0.75-0.90 were selected and fitted with a thermal bremsstrahlung model spectrum. The results for each phase interval are shown in Table 2. The column density stays constant at $\sim 5 \times 10^{20}\text{cm}^{-2}$ within error limits; whereas the temperature is high in the first three intervals of the orbital phase and minimum in the last. These results are not very reliable in view of the large error bars for the temperature, especially in the 0.25-0.40 orbital phase region where the count rate is low.

**Table 1.** Spectral Fits Parameters for the whole observation.

| Model | $N_H(10^{20}\text{cm}^{-2})$ | $A_i^a$ | kT(keV) | $\alpha$ | $(\chi_\nu^2)^b$ |
|---|---|---|---|---|---|
| Bremsstrahlung[1] | $4.68 \pm 0.20$ | $2.15 \pm 0.10 \times 10^{-3}$ | $5.09 \pm 1.42$ | | 2.0 |
| Raymond-Smith[2] | $4.31 \pm 0.28$ | $7.11 \pm 0.34 \times 10^{-3}$ | $3.93 \pm 0.74$ | | 1.6 |
| Power law[3] | $5.02 \pm 0.15$ | $2.13 \pm 0.16 \times 10^{-3}$ | | $1.48 \pm 0.07$ | 2.2 |
| Blackbody[4] | $0.89 \pm 0.19$ | $3.68 \pm 0.10 \times 10^{-4}$ | $0.35 \pm 0.01$ | | 1.5 |

$^a$ $A_i$ is the normalization coefficients of the different models denoted by $i = 1, 2, 3, 4$.
$^b$ Values of reduced $\chi_\nu^2$ for 77 degrees of freedom.
[1] $F(E) = A_1 e^{-E/kT}(E/E_o)^{-1} g(E; kT) e^{-N_H \sigma(E)}$; $A_1(photons/cm^2/s/keV)$ is the normalization constant, E is the photon energy in keV, $(E/E_o)$ is the photon energy normalized to 1 keV, T is the temperature in kelvins, k is the Boltzmann constant, g(E;kT) is the Gaunt factor (Kellogg et al. 1975), $N_H$ is the column density, $\sigma(E)$ is the photoelectric absorption cross section (Morrison and McCammon 1983).
[2] Emission spectrum from hot diffuse gas based on the Raymond and Smith (1977) calculations; $A_2(cm^{-5})$ is the normalization amplitude.
[3] $F(E) = A_3(E/E_o)^{-\alpha} e^{-N_H \sigma(E)}$; $A_3(photons/cm^2/s/keV)$, $\alpha$ is the power law index.
[4] $F(E) = A_4/T \frac{(E/T)^2}{e^{E/T}-1}$; $A_4(photons/cm^2/s)$ is the black body integral.

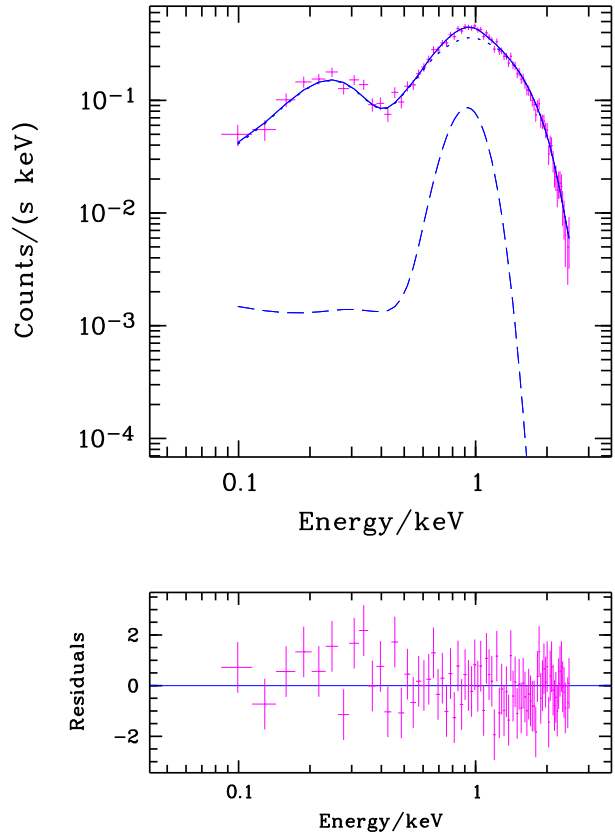

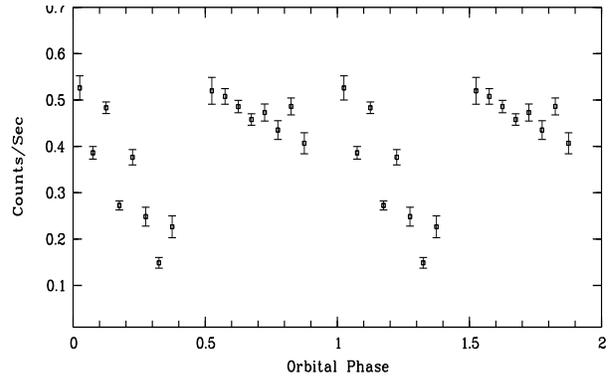

**Fig. 2.** The upper panel shows the spectrum of TT Ari as observed with the PSPC on board of Rosat, together with a bremsstrahlung spectrum at a temperature 5.09 keV modified with a 0.41 keV absorption edge and a 0.996 keV Gaussian emission line ($\chi_\nu^2 = 0.79$ per degrees of freedom). The emission line model is shown as a dashed line. The lower panel shows the residuals of the fit.

**Fig. 3.** The orbital modulation of TT Ari folded with the period and the spectroscopic epoch given by Thorstensen et. al. 1985.

**Table 2.** Spectral Fits Parameters for thermal bremsstrahlung model spectra for the selected phase intervals. Gaps due to ROSAT observation windows are excluded.

| Phases | 0.00-0.25 | 0.25-0.40 | 0.50-0.75 | 0.75-0.90 |
|---|---|---|---|---|
| $N_H^a$ | 5.5±0.4 | 5.1±1.1 | 4.6±0.3 | 4.7±0.4 |
| kT(keV) | 4.3±1.7 | 3.8±3.8 | 6.4±3.3 | 1.9±0.7 |
| $(\chi_\nu^2)^b$ | 3.09 | 2.18 | 3.2 | 1.21 |

[a] in units of ($10^{20}$cm$^{-2}$); [b] Values of reduced $\chi_\nu^2$ for 18 degrees of freedom.

### 1mHz Oscillations

The ROSAT observations are interrupted by a number of gaps which limit the timing analysis. In order to examine the broad shape of the power spectrum we performed a Fast Fourier Transform (FFT) for each observation window and averaged them. The length of each stretch is chosen to be 640 seconds to get high statistics covering most of the observation windows. The PDS have been normalized according to Leahy et al. (1983). The results are shown in Figure 4. White noise would give an average value of 2 in the power spectrum. It is clear from the figure that, there is no significant power excess at periods shorter than 100 seconds ($f \geq 10$mHz). From 100 seconds up to 1000 seconds (1mHz) there is excess power due to the satellite wobble, the observation windows and a red noise component from the source.

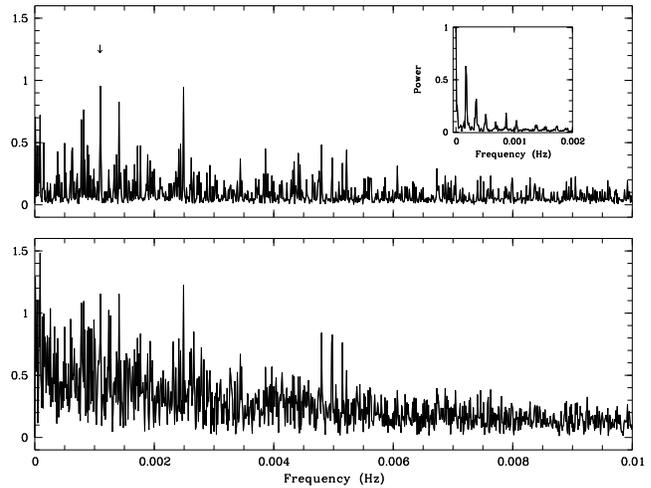

**Fig. 5.** The power spectrum of the TT Ari. The lower panel shows the discrete power spectra. The upper panel shows the spectrum after processing using CLEAN to remove artefacts due to window spectrum. The window spectrum is illustrated in the inset. The arrow indicate the 1.09 mHz oscillation (for the other powers see the text).

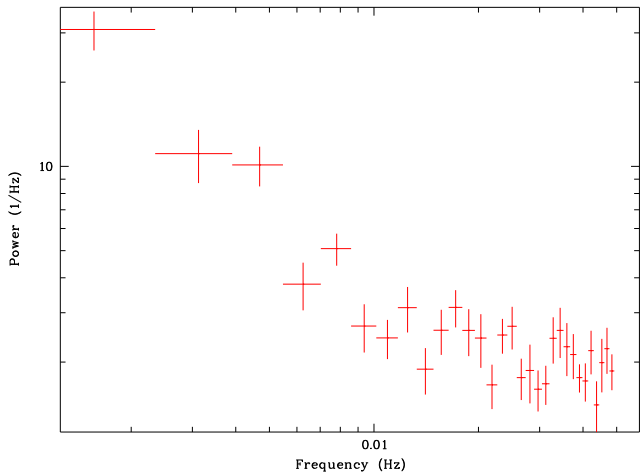

**Fig. 4.** The power spectrum of the TT Ari. Data were binned for 10 seconds and for each observation window the stretch lenghts of 640 seconds were averaged.

In order to search for the ∼1mHz oscillations in X-rays we estimated the discrete power spectrum (Deeming 1975) by correcting the photon arrival times to the solar system barycenter and integrating the data in bins of 30s. Since the data at hand is very unequally sampled we deconvolved the effects of the window function (or ghost powers) by using the "CLEAN" algorithm (Roberts et al. spectra in Figure 5. As can be seen in both power spectra the peak power values are at ∼1.09 mHz and 2.49 mHz. The first one corresponds to the previously known flickering oscillation (Jensen et al. 1983) and the second one is the wobble period of the PSPC detector which strongly dominates when the source is in the ring of ∼ 40$'$ at the center of PSPC. We also see peaks at ∼0.1 mHz (close to the orbital modulation frequency) ∼0.82 mHz, ∼0.87 mHz and ∼1.4 mHz. The 1.4 mHz oscillation can be identified as the beat frequency of the wobble period and the flickering oscillation (2.49 mHz - 1.09 mHz). It should be noted that the ∼1.09 mHz oscillation is consistent with optical oscillations recently found by Hollander and van Paradijs (1992). In order to see the changes of flickering oscillations we have divided the data into two equal segments. To see whether the luminosity is stable we have constructed the spectra and found that during the first half of the observation the source is ∼ 20% − 25% fainter than during the second half. The power spectra of the first and second halves are represented in Figure 6. In the power spectrum of the first half we see the flickering frequency at ∼ 1.09 ± 0.03mHz. In the second half the most dominant power (except the wobble period) appears at ∼ 1.06 ± 0.03mHz. The comparison of the power spectra of the two halves of the data indicates that when the source gets brighter the flickering frequency does not significantly change.

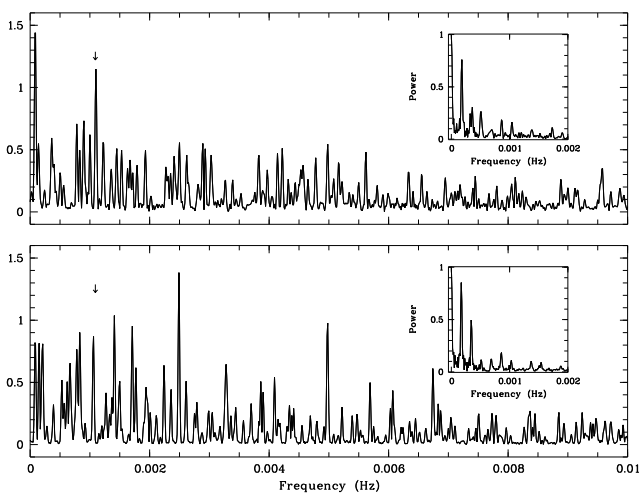

**Fig. 6.** The CLEANed power spectrum of the first and the second half of the observation, inset being the window spectrum. The arrow in the upper panel shows the 1.09 mHz oscillation. The arrow in the lower panel shows the 1.06 mHz oscillation.

## 4. Discussion

We have analysed the spectral and timing behaviour of TT Ari in the soft X-ray band using the archival ROSAT observations. The X-ray observation of TT Ari with the Einstein X-ray observatory measured a column density of $N_H \sim 15 \times 10^{20}$ cm$^{-2}$, a temperature of T $>10$ keV and a flux of $\sim 20 \times 10^{-12}$ erg cm$^{-2}$ s$^{-1}$ for a thermal bremsstrahlung fit between 0.2 and 4 keV (Jensen et al. 1983). Although our measured flux level is less than the Einstein value and the luminosity is smaller than the luminosity of $\sim 5.5 \times 10^{31}$ erg s$^{-1}$ obtained from combined ROSAT and GINGA observations in 0.2-37.4 keV range (Robinson and Cordova, 1993) for the same distance, these differences are probably due to the different energy ranges and the inadequacy of single temperature fits, so that the data is consistent with a constant luminosity. A 0.996 keV Gaussian emission line suggests the presence of iron L shell emission lines in thermal spectra (Belloni et al. 1991) and an absorption at 0.423 keV due to Ar IX or at 0.392 keV due to C V. The absorption feature may arise from calibration problems of the PSPC near the carbon edge around 0.28 keV, but it is also possible to observe such an effect if the X-rays are coming from near the hot spot and getting absorbed by cold matter in front. C V absorption is more likely because it has already been confirmed from the IUE observations that C IV is present (Robinson and Cordova, 1993).

Folding the data with the spectroscopic orbital period we noted a modulation in the count rate with a maximum at around orbital phases 0.5-0.9. We have ruled out the possibility that the orbital variation is an artefact dominated by a strong variation in a single orbit by following shown in Figure 1. $N_H$ remains constant, $\sim 5 \times 10^{20}$ cm$^{-2}$, through the orbit and bremsstrahlung temperature fits give a minimum temperature $1.9 \pm 0.7$ keV in the 0.75-0.90 phase interval. The temperature in other orbital phase intervals is higher though less accurately determined (Table 2). Van Teeseling and Verbunt (1994) have observed spectral variability with orbital phase in the CV GP Com. They find that the spectral and count rate variation in GP Com can be explained in terms of the orbital variability of the neutral hydrogen absorption column. For TT Ari the constancy of $N_H$ leads us to associate the variations with a soft X-ray source rather than variable absorption. Indeed, the minimum temperature is in the 0.75-0.90 orbital phase region, and this temperature fit has the smallest error bar. This coincides with high X-ray count rate. This source is most likely to be associated with a hot spot that is particularly visible around orbital phase 0.7-0.8 where the accretion stream from the secondary impacts on the outer rim of the accretion disk. The extended maximum in the count rate indicates an extended source around these binary phases, geometrically larger than the hot spot. Possible source of the emission could be a shock front or a corona lifted over the disk rim. Such a corona surrounding the hot spot might have its source in the mechanical energy deposited by the accretion column and dissipated in its impact on the disk. Having suggested that the increased $L_x$ at 0.75-0.90 orbital phase is associated with a soft X-ray source, we recall another observation of orbital variability peaking in the same phase interval of 0.75-0.90. Archival IUE data during and after the ROSAT observations show a similar variation on C IV ($\lambda 1549$): the equivalent width of the C IV absorption line (Robinson and Cordova, 1993) increases from a minimum at phase 0.2 to a maximum at 0.75, similar to the modulation of the total X-ray count rate. This increase in C IV absorption may also be associated with the hot spot in the outer rim of the accretion disk, around 0.75 orbital phase. The C IV absorption probably does not occur at the same soft X-ray emission although both are associated with the hot spot. For example the hot spot itself may be at the appropriate temperature for C IV absorption, and the soft X-ray emitting corona may extend above and below the disk. The minimum luminosity, possibly harder, X-rays and minimum C IV absorption occur in the 0.0-0.25 orbital phase interval where the hot spot and associated corona are out of the line of sight and may even be partially eclipsed, depending on the inclination.

Timing analysis indicates that while the source gets brighter, the frequency of the 1mHz oscillation does not change with the X-ray intensity. However, if the beat frequency model (Alpar and Shaham 1985) holds, a frequency change of more than 0.1 mHz may be expected, depending on the parameters. In this model the quasi periodic oscillation frequency is the beat frequency between the Kepler frequency at the inner edge of the ac-

$\nu_{QPO} = \nu_K - \nu \approx a\dot{M}^{3/7} - \nu_*$, where $a$ is a constant involving the star's magnetic field, and the QPO frequency increases with increasing mass accretion rate. A variation in luminosity would lead to $\Delta\nu_{QPO}/\nu_{QPO} = (3/7)(1+\nu_*/\nu_{QPO})(\Delta L/L)$ which gives $\Delta\nu_{QPO} > 0.1$mHz for $(\Delta L/L) \sim 0.25$ as observed, if $L \propto \dot{M}$. However $\nu_{QPO}$ should be less if $L \propto \dot{M}^\beta$, $\beta < 1$. Thus the observation that $\nu_{QPO}$ remains constant with $(\Delta L/L) \sim 0.25$ changes in luminosity may still be compatible with the beat frequency model. The beat frequency model explains the short time scale correlated changes in the intensity and the QPO frequency in low mass X-ray binaries in their horizontal branch state. The application in TT Ari has to do with long time scale ($\sim 30yr$) secular changes in the optical luminosity. It is interesting to note that while the optical flickering (or quasi-periodic) frequency is correlated with optical luminosity on these long time scales in agreement with the beat frequency model (Hollander and van Paradijs 1992), the X-ray flickering frequency does not show the same property on short time scales, at least for the model parameter $\beta = 1$. This is not surprising as it is the optical luminosity in this source that is directly related to accretion and the beat frequency. In TT Ari the X-ray luminosity is much lower than the optical, $L_x/L_{opt} \sim 0.1$, and some of the X-rays possibly come from a corona above the disk. The X-ray luminosity changes may not correlate with those in the optical. To check the validity of the beat frequency model, the frequency intensity relation in the optical must be tested on short time scales. It will be interesting to investigate wheather the variation in the X-ray and optical luminosities are correlated. This can be achieved if the source is monitored simultaneously in X-ray and optical bands.

*Acknowledgements.* We thank the refree, Jan van Paradijs, for a careful reading and valuable comments, and Keith Horne and André van Teeseling for a useful discussion. This work is supported by The Scientific and Technical Research Council of Turkey, TUBITAK, under High Energy Astrophysics Unit Project TBAG-Ü/18-2. E.N. Ercan and G. İkis would like to thank Turkish-Balkan Physics Research and Application Center of Boğaziçi University for financial support. M.A.Alpar thanks the Academy of Science of Turkey for supporting his research.